\def\kB{k_{\text B}}

\def\kF{k_{\text{F}}}
\def\vF{v_{\text{F}}}
\def\NF{N_{\text{F}}}

\def\epsilonF{\epsilon_{\text F}}

\def\be{\begin{equation}}
\def\ee{\end{equation}}
\def\bea{\begin{eqnarray}}
\def\eea{\end{eqnarray}}
\def\bse{\begin{subequations}}
\def\bSe{\begin{SEquation}}
\def\ese{\end{subequations}}\def\eSe{\end{SEquation}}

\def\DT{D_{\text{T}}}
\def\perpgradT{({\hat{\bm k}}_{\perp}\!\!\cdot\!{\bm\nabla}T)}
\def\Ssym{S^{\text{sym}}}
\def\gradperpT{\nabla_{\!\perp}T}

\newcounter{Sequ}
\newenvironment{SEquation}
  {\stepcounter{Sequ}%
    \addtocounter{equation}{-1}%
    \equation}
  {\endequation}%
\documentclass[prl,twocolumn,amsmath,amssymb,preprintnumbers]{revtex4}
\usepackage{graphicx}
\usepackage{dcolumn}
\usepackage{bm}
\usepackage{bbm}
\usepackage{color}
\usepackage{soul}
\begin{document}

\bibliographystyle{unsrtnat}

\title{A fluctuation-dissipation relation in a non-equilibrium quantum fluid}

\author{T.R. Kirkpatrick$^{1}$ and D. Belitz$^{2,3}$}

\affiliation{$^{1}$Institute for Physical Science and Technology, University of Maryland, College Park, MD 20742, USA\\
 $^{2}$Department of Physics and Institute for Fundamental Science, University of Oregon, Eugene, OR 97403, USA \\
 $^{3}$ Materials Science Institute, University of Oregon, Eugene, OR 97403, USA}

\date{\today}
\begin{abstract}

{There is no simple fluctuation-dissipation theorem (FDT) for nonequilibrium systems. We show that for a fluid in a non-equilibrium 
steady state (NESS) characterized by a constant temperature gradient there is a generalized FDT that relates commutator 
correlation functions to the bilinear response of products of observables. This allows for experimental probes of the long-range 
correlations in such a system, quantum or classical, via response experiments. We also show that the correlations are not tied 
to thermal fluctuations but are intrinsic to the NESS and reflect a generalized rigidity.}

 \end{abstract}

\maketitle

A fundamental result of equilibrium statistical mechanics is the fluctuation-dissipation theorem, which states that the system's
linear response to an external perturbation is related to the fluctuations in the equilibrium state \cite{Nyquist_1928, Callen_Welton_1951}.
Specifically, the response functions are proportional to the corresponding correlation functions, and in a classical system the
proportionality factor is simply the inverse temperature \cite{Forster_1975}. In a non-equilibrium system this is no longer true.
{There is a substantial body of work on the properties of fluctuations far from equilibrium (see, e.g., Refs.~\onlinecite{Sevick_et_al_2008},
\onlinecite{Gaspard_2022} and references therein), on linear response in non-equilibrium systems \cite{Baiesi_Maes_2013},
and on aspects of relations between the two \cite{Andrieux_Gaspard_2004, Maes_2020, Gaspard_2022}, but there is no general prescription for
probing fluctuations via the response to external perturbations.}
{Consequently, fluctuations can be observed only via scattering experiments, which become increasingly difficult as the
temperature is lowered, but not via the linear response to a macroscopic perturbation.}
One of the main results reported in this Letter is that for a fluid,
classical or quantum, in a non-equilibrium steady state (NESS) characterized by a fixed temperature gradient there still is a
relation between fluctuations and response, but it is not linear. Rather, the {non-equilibrium parts of the} correlation functions 
{determine} 
the bilinear response of products of observables to external perturbations, {as we will demonstrate in Eqs.~(\ref{eq:13}) 
and (\ref{eq:17}) below.} Since perturbations can be controlled experimentally
independent of other parameters, this opens new avenues for observing correlations, especially in
quantum fluids, where thermal fluctuations are weak due to the low temperature. In addition, we elucidate several other
aspects of fluids in such a NESS, especially in the quantum limit. 

To motivate our investigations we recall that classical simple fluids subject to a fixed temperature gradient harbor correlations 
that are extraordinarily long-ranged \cite{Kirkpatrick_Cohen_Dorfman_1982c, Dorfman_Kirkpatrick_Sengers_1994, Ortiz_Sengers_2007}.
For instance, the equal-time temperature-temperature correlation function diverges for small wave numbers $k$ as
$1/k^4$. In real space, this corresponds to correlations that extend over the entire width of the sample and decay
on the same scale \cite{Ortiz_Sengers_2007}. These correlations 
are generic in the sense that they do not require any fine tuning of parameters, and they are not related to any broken
symmetry. Rather, they are the result of the coupling of the temperature fluctuations
to the diffusive shear velocity. This surprising result has been confirmed theoretically by means of a variety of
techniques \cite{Kirkpatrick_Cohen_Dorfman_1982a, Ronis_Procaccia_1982, Ortiz_Sengers_2007}, and it has
been observed by many light scattering experiments, see Ref.~\cite{Sengers_Ortiz_Kirkpatrick_2016} and references therein.

Despite being well established and confirmed, this phenomenon raises several questions that historically have
not been emphasized. One is the fact, mentioned above, that the relevant correlation functions in a NESS are not 
in any obvious way related to response functions. Another question is whether
the  long-ranged correlations are tied to thermal fluctuation effects, or whether they are
more generic and reflect some type of generalized rigidity \cite{Anderson_1984} that is present even in the zero-temperature limit
and also manifests itself in the response of the system to external perturbations. 
Recent work on classical fluids has suggested the latter \cite{Kirkpatrick_Belitz_Dorfman_2021}, but a relation
between correlation functions and response theory has been lacking. {Part of the purpose of this Letter is
to provide such a relation.}

{The missing correlation-response relation discussed above is equally relevant for classical
and quantum {fluids}, but it poses a particularly significant problem for the latter {\cite{quantum_hydrodynamics_footnote}}. Direct}
measurements of the correlation functions via light scattering are difficult even in 
the classical case because of the very small scattering angles required. With decreasing temperature the 
fluctuations become smaller, which makes it even more desirable to observe the effect via response experiments, 
if feasible. Specifically, 
there are two obvious types of correlation functions: symmetrized, or anticommutator correlation functions
{that we denote by $S^{\text{sym}}$,}
and anti-symmetrized, or commutator ones {that we denote by $\chi''$ (this  is the customary notation
for the commutator correlation function \cite{Forster_1975}, with the double prime indicating that this is the spectrum, 
{or spectral density,} of a causal function).}
As functions of the wave
vector ${\bm k}$ and the frequency $\omega$ they are defined by {\cite{Forster_1975, classical_limit_footnote}}
\begin{widetext}
\bse
\label{eqs:1}
\bea
\frac{1}{2}\,\left\langle\left[\delta {\hat A}({\bm k}_1,\omega_1), \delta {\hat B}({\bm k}_2,\omega_2)\right]_{+}\right\rangle
&=& V \delta_{{\bm k}_1,-{\bm k}_2}\,2\pi \delta(\omega_1+\omega_2)\,\Ssym_{AB}({\bm k}_1,\omega_1) \ ,
\label{eq:1a}\\
\frac{1}{2\hbar}\,\left\langle\left[\delta {\hat A}({\bm k}_1,\omega_1), \delta {\hat B}({\bm k}_2,\omega_2)\right]_{-}\right\rangle
&=& V \delta_{{\bm k}_1,-{\bm k}_2}\,2\pi \delta(\omega_1+\omega_2)\, \chi_{AB}''({\bm k}_1,\omega_1) \ .
\label{eq:1b}
\eea
\ese
\end{widetext}
Here $\delta\hat A$, $\delta\hat B$ are operator-valued fluctuations of observables, $[\ ,\ ]_+$ and $[\ ,\ ]_-$ denote
anticommutators and commutators, respectively, $\langle \ldots\rangle$ denotes a quantum mechanical expectation 
value plus {a statistical mechanics} average, and $V$ is the system volume. {We use carets to denote operator-valued quantities
(and below also to denote unit vectors; this should not lead to confusion); the corresponding quantities without carets
are number-valued classical objects. The two types of correlation functions}
are related by 
\be
\Ssym_{AB}({\bm k},\omega) = \chi''_{AB}({\bm k},\omega)\,{\hbar}\coth({\hbar}\,\omega/2T)\ .
\label{eq:2}
\ee
{Here, and throughout the paper, we put $\kB=1$, i.e., we measure the temperature in units of energy.
We note that the exact relation between $\Ssym$ and $\chi''$ is more complicated in systems that are not
spatially homogeneous. It reduces to Eq.~(\ref{eq:2}) if the local temperature is replaced by its spatial average.}
The corresponding static correlation functions are
\bse
\label{eqs:3}
\bea
\Ssym_{AB}({\bm k}) &=&  \int_{-\infty}^{\infty} \frac{d\omega}{2\pi}\,\Ssym_{AB}({\bm k},\omega)\ ,
\label{eq:3a}\\
\chi_{AB}({\bm k}) &=& \int_{-\infty}^{\infty} \frac{d\omega}{\pi}\,\chi_{AB}''({\bm k},\omega)/\omega\ .
\label{eq:3b}
\eea
\ese
The symmetrized correlation functions $\Ssym_{AB}$ are observable by means of scattering experiments \cite{Forster_1975}.
The physical meaning of the antisymmetrized correlation functions $\chi''_{AB}$ is {\em a priori} less obvious.
In an equilibrium system, where the correlations are generically short-ranged, they describe the linear response of the 
system to external fields{, as follows. Let $h_B$ be the external field conjugate to ${\hat B}$. Then to linear
order in the fields one has \cite{Forster_1975} }
\be
{\delta\langle{\hat A}\rangle({\bm k},\omega) = \chi_{AB}({\bm k},\omega)\,h_B({\bm k},\omega)\ ,}
\label{eq:3.1}
\ee
{where $\chi_{AB}({\bm k},\omega) = \int_{-\infty}^{\infty} dx\,\chi_{AB}''({\bm k},x)/(x - \omega - i0)$
with $i0$ an infinitesimal imaginary part.}
That is, the equilibrium fluctuations determine the linear response, which to second order in the external
field yields the energy dissipated by the system. This is the content of the fluctuation-dissipation theorem 
\cite{Nyquist_1928, Callen_Welton_1951}. In a NESS, the 
relation (\ref{eq:2}) still holds
{(if the local temperature is replaced by its spatial average $T$)}, but the
commutator correlations functions no longer describe the linear response and the usual fluctuation-dissipation
theorem breaks down. 

In this Letter we identify the quantum analogs of the classical long-ranged correlations. Our two main results
are: (1) The long-ranged commutator  correlation functions are still related to response functions, but the 
relation is not a simple proportionality. Rather, the non-equilibrium contributions to the correlation functions 
are related to the {bilinear} response of products of observables, {see Eqs.~(\ref{eq:13}) and (\ref{eq:17}) below.}
(2) In a modified form, the long-ranged correlations extend to zero temperature. This shows that they
are not tied to thermal fluctuations, although thermal fluctuations can be used to probe them. Rather, they
are an inherent long-wavelength property of the NESS and indeed reflect a novel type of generalized
rigidity that does not become weaker with decreasing temperature. 
We will start by explaining the second result, and then demonstrate the first one. 

\begin{figure}[b]
\includegraphics[width=8cm]{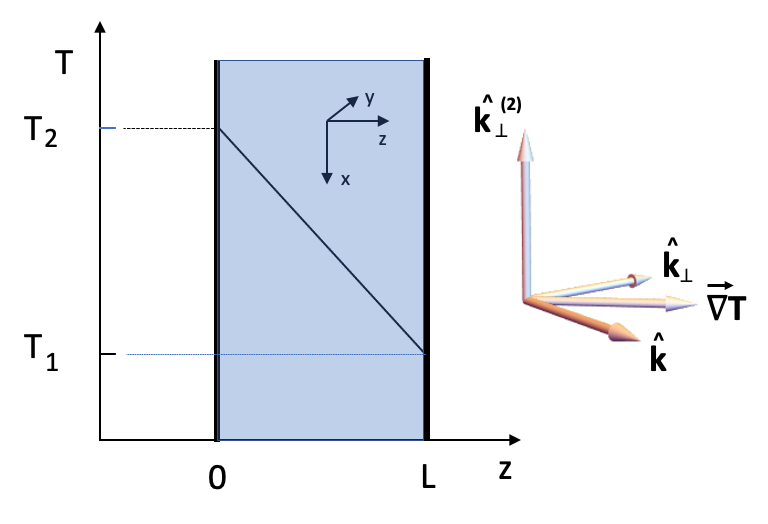}
\caption{A fluid subject to a constant temperature gradient in the $z$-direction between two parallel confining plates.
              {$\hat{\bm k}$, $\hat{\bm k}_{\perp}$, and $\hat{\bm k}_{\perp}^{(2)}$ are unit vectors that span the
              wave-number space, with the coordinate system chosen such that $\hat{\bm k}$ and $\hat{\bm k}_{\perp}$ 
              are coplanar with the temperature gradient.}}
\label{fig:1}
\end{figure}
\begin{figure}[t]
\includegraphics[width=8cm]{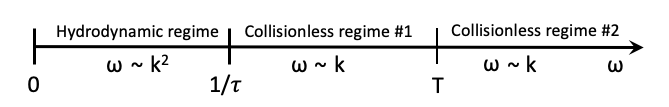}
\caption{Frequency regimes. {The hydrodynamic regime is characterized by a diffusive shear velocity, with the frequency
              $\omega$ scaling as the wave number squared. In the collisionless regimes the shear velocity is a propagating mode
              with the frequency scaling linearly with the wave number.}}
\label{fig:2}
\end{figure}

Consider a fluid confined between two plates that is subject to a constant temperature gradient ${\bm\nabla}T$
in the $z$-direction, see Fig.~\ref{fig:1}. Let ${\bm k}$ be the wave vector of a temperature fluctuation, and let the wave-vector space be
spanned by three mutually perpendicular unit vectors $\hat{\bm k} = {\bm k}/k$, $\hat{\bm k}_{\perp}$, and $\hat{\bm k}_{\perp}^{(2)}$
such that $\hat{\bm k}_{\perp}$ is coplanar with ${\bm k}$ and ${\bm\nabla}T$. For our
purposes the temperature gradient will appear in the combination $\nabla_{\!\perp}T \equiv \hat{\bm k}_{\perp}\!\cdot\!{\bm\nabla} T$.
For definiteness, we consider a fermionic quantum fluid (e.g., conduction electrons in metals). {Analogous
effects must be present in bosonic fluids as well, but at asymptotically low temperatures Bose-Einstein condensation will lead to 
complications that require additional investigation.}
Let $\tau$ be the relaxation time, $\vF$ the Fermi velocity,
and $\omega$ the frequency. We need to distinguish between the hydrodynamic regime, where $\omega < 1/\tau$ 
or $\vF k < 1/\tau$, and the collisionless regime where $\omega > 1/\tau$. The latter is in general subdivided
into regimes where ${\hbar}\,\omega < T$ and ${\hbar}\,\omega > T$, respectively, see Fig.~\ref{fig:2} \cite{Planckian_footnote}.
{Here $T$ is the spatially averaged temperature {\cite{spatial_averaging_footnote}}.}
We are interested in the correlation functions for the temperature $T$ and the component 
$u_{\perp} \equiv \hat{\bm k}_{\perp}\cdot{\bm u}_{\perp}$ of the fluid shear velocity ${\bm u}_{\perp}$.
{The nature of the latter changes from diffusive to propagating as one goes from the hydrodynamic regime
to the collisionless one.
In the latter a hydrodynamic description no
longer applies and one has to work with a quantum kinetic theory.}
In what follows we first state our results and then comment on their derivations. 
{For details see the Supplemental Material \cite{SM_footnote}.}

\noindent
{\it Hydrodynamic Regime:} In the hydrodynamic regime {one has $\omega < 1/\tau < T/\hbar$ \cite{Planckian_footnote},
and the correlation functions are the same as in a classical fluid. In particular, $S_{AB}^{\text{sym}}$ becomes a classical
correlation function $\langle A\,B\rangle$ with the angular brackets denoting a classical {statistical mechanics} average, and we have }
\cite{Kirkpatrick_Cohen_Dorfman_1982c}:
\bse
\label{eqs:4}
\bea
\Ssym_{TT}({\bm k},\omega) &=& \frac{2 T^2}{c_V}\,\frac{\DT k^2}{\omega^2 + (\DT k^2)^2}
\nonumber\\
&& \hskip -40pt +\, (\gradperpT)^2\,\frac{2T}{\rho}\,\frac{\nu k^2}{(\omega^2 + (\DT k^2)^2)(\omega^2 + (\nu k^2)^2)} \qquad
\label{eq:4a}
\eea
The first term is the standard equilibrium contribution,
and the second term is the long-ranged NESS contribution. $c_V$ and $\rho$ are the spatially averaged specific heat 
per volume and the fluid {mass} density, respectively, and $\DT$ and $\nu$ are the spatially averaged heat diffusion coefficient
and kinematic viscosity, respectively. The velocity-velocity correlation function is the same as in equilibrium,
\be
\Ssym_{u_{\perp}u_{\perp}}({\bm k},\omega) = \frac{2T}{\rho}\,\frac{\nu k^2}{\omega^2 + (\nu k^2)^2}\ ,
\label{eq:4b}
\ee
and the mixed symmetrized correlation functions are
\bea
\Ssym_{u_{\perp}T}({\bm k},\omega) &=& \Ssym_{T u_{\perp}}(-{\bm k},-\omega)
\nonumber\\
&=&  \gradperpT\,\frac{1}{\rho}\, \frac{\nu k^2}{\omega^2 + (\nu k^2)^2}\,\frac{2T}{i\omega + \DT k^2}\ .
\label{eq:4c}\
\eea
The corresponding commutator correlation functions are {given by Eq.~(\ref{eq:2}) with $\hbar \coth(\hbar\,\omega/2T) \approx 2T/\omega$:}
\be
\chi''_{AB}({\bm k},\omega) = \Ssym_{AB}({\bm k},\omega)\,\omega/2T \qquad (A,B = T, u_{\perp})\ .
\label{eq:4d}
\ee
\ese 
The static correlation functions are
\bse
\label{eqs:5}
\bea
\Ssym_{TT}({\bm k}) &=&  \frac{T^2}{c_V} + \frac{T (\gradperpT)^2}{\rho \DT (\nu + \DT) \, k^4}\ ,
\label{eq:5a}\\
\Ssym_{u_{\perp}u_{\perp}}({\bm k}) &=&T/\rho\ ,
\label{eq:5b}\\
\Ssym_{u_{\perp}T}({\bm k}) &=& \Ssym_{T u_{\perp}}(-{\bm k}) = \gradperpT\, \frac{T/\rho}{(\nu + \DT)k^2}\ ,
\label{eq:5c}
\eea
and 
\be
\chi_{AB}({\bm k}) = \Ssym_{AB}({\bm k})/T \qquad (A,B = T, u_{\perp})\ .
\label{eq:5d}
\ee
\ese
{$\Ssym_{TT}({\bm k})$ in Eq.~(\ref{eq:5a}) is the equal-time temperature-temperature correlation function mentioned in the
introduction that diverges as $1/k^4$. This result is exactly the same as for a classical fluid, as expected in the
hydrodynamic regime.}

\medskip
\noindent
{\it Collisionless Regime:} In the collisionless regime the expressions for the dynamic correlation functions
are lengthy {and are given in the Supplemental Material, together with an outline of their derivations \cite{SM_footnote}.}
Here we give only the results for the static temperature-temperature correlations. To leading
order as $T\to 0$ we have
\be
\chi_{TT}({\bm k}) = \frac{1}{\NF}\,\frac{3}{\pi^2} \left[1 + \frac{\pi^2}{12}\,(2\pi^2-3)\,\frac{(\gradperpT)^2}{\epsilonF^2 k^2}\right]\ 
\label{eq:6}
\ee
everywhere in the collisionless regime. {Here $\NF$ is the density of states at the Fermi level, and $\epsilonF$ is the Fermi energy.}
The symmetrized correlation function is $\Ssym_{TT}({\bm k}) = T\chi_{TT}({\bm k})$
in {Collisionless} Regime \#1 in Fig.~\ref{fig:2} and, apart from a factor of $O(1)$, $\Ssym_{TT}({\bm k}) \approx \vF k\,\chi_{TT}({\bm k})$
in {Collisionless} Regime \#2.

To derive these results we note that in the hydrodynamic regime the usual Navier-Stokes equations are applicable, 
as they are based on very general physical principles, {viz., the conservation laws for mass, momentum, and
energy combined with force-balance considerations \cite{Landau_Lifshitz_VI_1987, Chaikin_Lubensky_1995, us_fluctuating_hydrodynamics}.
Consequently, they}
hold for quantum fluids as well as for classical ones. In order 
to calculate correlation functions, they need to be augmented by fluctuating forces \cite{Landau_Lifshitz_VI_1987,
Kirkpatrick_Belitz_2022}. Of the various nonlinearities we need to keep only the crucial coupling between the
temperature fluctuation and the transverse velocity, which turns into a linear term in the presence of a fixed
temperature gradient. We can further ignore pressure fluctuations, which lead to sound waves (or plasmons in a
charged Fermi liquid) that are much
faster than the diffusive shear fluctuations. The equations for the operator-valued temperature and shear
velocity then are
\bse
\label{eqs:7}
\be
\left(-i\omega + \nu {\bm k}^2\right) {\hat{u}}_{\perp}({\bm k},\omega) = {\hat P}_{\perp}({\bm k},\omega)\ ,\qquad\quad
\label{eq:7a}
\ee
\vskip -20pt
\be
\left(-i\omega + D_T {\bm k}^2 \right) {\hat T}({\bm k},\omega) + (\gradperpT) {\hat{u}}_{\perp}({\bm k},\omega) =  
{\hat Q}({\bm k},\omega)\ .
\label{eq:7b}
\ee
\ese
Here the transport coefficients $\nu$ and $\DT$ are to be understood as having been spatially averaged.
${\hat P}_{\perp}$ and $\hat Q$ are operator-valued fluctuating forces that are Gaussian distributed with zero mean. 
Their second moments can be obtained from the somewhat more general expressions derived in Ref.~\cite{Kirkpatrick_Belitz_2022}
{and are given explicitly in the Supplemental Material \cite{SM_footnote}.}
The $u_{\perp}$-$u_{\perp}$ correlation then is determined by the $P_{\perp}$ correlation and given
by Eq.~(\ref{eq:4b}). This is the same as in equilibrium since the temperature does not couple into the
$u_{\perp}$ equation. The equilibrium part of the $T$-$T$ correlation is given by the $Q$ correlation,
whereas the non-equilibrium part, as well as the mixed $T$-$u_{\perp}$ correlation, is given in terms of the
$u_{\perp}$ fluctuation, with $\gradperpT$ serving as a coupling constant. This yields the second term
in Eq.~(\ref{eq:4a}), and Eq.~(\ref{eq:4c}). The leading singularity in the static $T$-$T$ correlation function
is $1/k^4$, and $1/k^2$ in the $T$-$u_{\perp}$ correlation function, as in a classical fluid. This is because
the hydrodynamic equations are the same in either case.

In order to properly describe the collisionless regime one has to work with the fluctuating quantum kinetic theory 
developed in Ref.~\cite{Kirkpatrick_Belitz_2022}, 
{see the Supplemental Material for a derivation \cite{SM_footnote}.}
However, the qualitative features of the results can be obtained from simple scaling arguments as follows.
In the collisionless regime the diffusive inverse propagators of the form ${\cal D}^{-1}({\bm k},\omega) = \omega + iDk^2$
that appear on the left-hand sides of Eqs.~(\ref{eqs:7}), where $D$ can represent either $\nu$ or $\DT$, effectively 
turn into propagating zero modes of the form $\omega \mp ck$, with $c \approx \vF$ the propagation speed.
The transport coefficients thus effectively become singular functions of the wave number and scale as $D \sim \vF/k$.
The low-temperature result for $\chi_{TT}$, Eq.~(\ref{eq:6}), then follows from Eq.~(\ref{eq:5a}) by replacing
$\nu, \DT \to \vF/k$ and using the low-temperature expression for the specific heat, {$c_V \propto \NF T$} (prefactors, as
well as issues regarding reality and signs, require a more detailed analysis).
For the symmetrized correlation function one needs to consider the frequency integral in Eq.~(\ref{eq:3a}) and
recognize that for the equilibrium contribution it needs to be cut off at $\omega \approx \DT k^2$. In the limits
$\DT k^2 \ll T{/\hbar}$ and $\DT k^2 \gg T{/\hbar}$, and using again the effective scaling of $\DT$ and $\nu$ explained above,
one then obtains the relations between $\Ssym_{TT}({\bm k})$ and $\chi_{TT}({\bm k})$ given after Eq.~(\ref{eq:6}).

As mentioned above, the commutator functions $\chi''$ do {\em not} determine the linear response of the
NESS to external perturbations, in contrast to an equilibrium system. In order to determine the response
functions we consider the averaged Navier-Stokes equations in the presence of a field $h_{u_{\perp}}$
conjugate to 
$u_{\perp}$:
\bse
\label{eqs:8}
\bea
\left(-i\omega + \nu k^2\right) u_{\perp}({\bm k},\omega) &=& \frac{\nu}{\rho}\, k^2 h_{u_{\perp}}({\bm k},\omega)\ ,
\label{eq:8a}\\
\left(-i\omega + D_T k^2\right) \delta T({\bm k},\omega) &+& (\gradperpT) u_{\perp}({\bm k},\omega) = 
\nonumber\\
&& \hskip -20pt \frac{1}{\rho}(\gradperpT) h_{u_{\perp}}({\bm k},\omega)\  . \qquad
\label{eq:8b}
\eea
\ese
{To avoid misunderstandings, we stress that these are equations for averaged, classical fluctuations
$\delta T$ and $u_{\perp}$. They are the standard Navier-Stokes equations \cite{Landau_Lifshitz_VI_1987}
except that the sound modes have been omitted since they occur on time scales much faster than the
diffusive shear and temperature fluctuations. They are driven by an external field $h_{u_{\perp}}$ that
essentially represents a shift of the velocity reference frame and can be regarded as a field conjugate
to $u_{\perp}$ \cite{Hohenberg_Halperin_1977, Chaikin_Lubensky_1995}. It}
can be experimentally realized by imposing a shear velocity on the system. {For a derivation from kinetic
theory, see Ref.~\onlinecite{Kirkpatrick_Belitz_2023a}.} The response functions
$X_{AB}({\bm k},\omega)$ are defined by
\bse
\label{eqs:9}
\bea
u_{\perp}({\bm k},\omega) &=& X_{u_{\perp}u_{\perp}}({\bm k},\omega)\,h_{u_{\perp}}({\bm k},\omega) \ .
\label{eq:9a}\\
\delta T({\bm k},\omega) &=& X_{Tu_{\perp}}({\bm k},\omega)\,h_{u_{\perp}}({\bm k},\omega) \ ,
\label{eq:9b}
\eea
\ese
From Eqs.~(\ref{eqs:8}) we find
\bse
\label{eqs:10}
\bea
X_{u_{\perp}u_{\perp}}({\bm k},\omega) &=& \frac{1}{\rho}\,\frac{\nu k^2}{-i\omega + \nu k^2}\ .
\label{eq:10a}\\
X_{Tu_{\perp}}({\bm k},\omega) &=& \frac{1}{\rho}\,(\gradperpT)\,\frac{1}{-i\omega + \DT k^2}\,\frac{-i\omega}{-i\omega + \nu k^2}\ .
\nonumber\\
\label{eq:10b}
\eea
\ese
From Eqs.~(\ref{eq:4b}) and (\ref{eq:10a}) we see that the spectrum {\cite{semantics_footnote}} of 
$X_{u_{\perp}u_{\perp}}$, $X''_{u_{\perp}u_{\perp}}({\bm k},\omega) = \text{Im}\,X_{u_{\perp}u_{\perp}}({\bm k},\omega)$,
equals the commutator correlation function $\chi''({\bm k},\omega)$, as expected. However, the spectrum of
$X_{Tu_{\perp}}$,
\be
X_{Tu_{\perp}}''({\bm k},\omega) = -(\gradperpT)\,\frac{\omega(\nu \DT k^2 - \omega^2)}{(\omega^2 + \DT^2 k^4)(\omega^2 + \nu^2 k^4)}\ ,
\label{eq:11}
\ee
while showing the same scaling behavior as $\chi''_{Tu_{\perp}}$, see Eqs.~(\ref{eq:4c}, \ref{eq:4d}), is not identical with the latter.
In particular, the static response function vanishes,
\be
X_{Tu_{\perp}}({\bm k}) = \int \frac{d\omega}{\pi}\,\frac{X_{Tu_{\perp}}''({\bm k},\omega)}{\omega} = 0\ ,
\label{eq:12}
\ee
while the static correlation function is nonzero, see Eq.~(\ref{eq:5c}). Nonetheless, the response functions still
provide a way to directly measure the commutator correlation functions, without relying on their symmetrized
counterparts that are suppressed at low temperatures. Specifically, considering Eqs.~(\ref{eqs:9}), (\ref{eqs:10}), and
(\ref{eq:4c}, \ref{eq:4d}) we have
\be
\delta T({\bm k},\omega)\,u_{\perp}(-{\bm k},-\omega) = \frac{i}{\rho}\,\chi''_{Tu_{\perp}}({\bm k},\omega)\,\vert h_{u_{\perp}}({\bm k},\omega)\vert^2
\label{eq:13}
\ee
That is, the commutator $T$-$u_{\perp}$ correlation function describes the bilinear response of the product of the
temperature and the shear velocity to the field conjugate to $u_{\perp}$. 
Similarly, the non-equilibrium part of the commutator $T$-$T$ correlation function can be expressed as a bilinear
response to the field $h_{u_{\perp}}$. We define an observable
\be
{\tilde T}({\bm k},\omega) = T({\bm k},\omega) - \frac{1}{\rho}\,(\gradperpT)\,\frac{1}{-i\omega + \DT k^2}\,h_{u_{\perp}}({\bm k},\omega)
\label{eq:14}
\ee
that obeys the equation
\be
\left(-i\omega + D_T k^2\right) \delta {\tilde T}({\bm k},\omega) + (\gradperpT) u_{\perp}({\bm k},\omega) = 0\ .
\label{eq:15}
\ee
The physical interpretation of Eq.~(\ref{eq:15}) is the heat equation with a streaming term that contains the
absolute shear velocity, whereas Eq.~(\ref{eq:8b}) contains the shear velocity relative to the field $h_{u_{\perp}}$.
The response of ${\tilde T}$ to the field $h_{u_{\perp}}$ is given by a response function
\be
X_{{\tilde T}u_{\perp}}({\bm k},\omega) = \frac{1}{\rho}\,(\gradperpT)\,\frac{-1}{-i\omega + \DT k^2}\,\frac{\nu k^2}{-i\omega + \nu k^2}\ .
\label{eq:16}
\ee
From Eqs.~(\ref{eq:16}), (\ref{eq:10b}), (\ref{eq:4a}), and (\ref{eq:4d}) we find that the non-equilibrium part of the commutator 
$T$-$T$ correlation function describes the bilinear response of the product of $\tilde T$ and $T$ to the field $h_{u_{\perp}}$:
\be
\delta{\tilde T}({\bm k},\omega)\,\delta T (-{\bm k},-\omega) = \frac{i}{\rho}\,\chi_{TT}''^{\text{\,neq}}({\bm k},\omega)\,\vert h_{u_{\perp}}({\bm k},\omega)\vert^2
\label{eq:17}
\ee

The Eqs.~(\ref{eq:13}) and (\ref{eq:17}) constitute our main result. They demonstrate that in a NESS characterized
by a constant temperature gradient the commutator correlation functions for the temperature and the shear
velocity are still related to the response of the system to an external shear perturbation, even though the usual
fluctuation-dissipation theorem is not valid. In contrast to the situation in equilibrium, where the correlation
functions are the same as the response functions, in a NESS the correlation functions are given by the
bilinear response of products of observables. {We note that Eqs.~(\ref{eq:13}) and (\ref{eq:17}) involve the
non-equilibrium parts of the correlation functions only. For $\Delta T \to 0$, $\chi''_{T u_{\perp}}$ vanishes and
$\chi''_{TT}$ reduces to its equilibrium part that obeys the usual fluctuation-dissipation theorem.}

To summarize, we have established a relation between correlation functions and response functions for a fluid in
a NESS. The resulting NESS fluctuation formulas (\ref{eq:13}) and (\ref{eq:17}) resemble the
equilibrium fluctuation-dissipation theorem, Eq.~(\ref{eq:2}) in the hydrodynamic limit, with the symmetrized
correlation function replaced by the product of two averaged observables, and the temperature replaced by
the driving field squared. The driving field, which is realized by an imposed shear velocity, one has experimental 
control over regardless of how low the temperature is. 
{We stress that we have derived this relation between the fluctuation functions and the response functions
only for the special case of a constant temperature gradient. Their structure suggests that they might hold for
more general nonequilibrium states as well, but whether or not that is true is an open question.}
We also have determined the quantum analogs of the
long-ranged correlations known to exist in a classical fluid in a NESS. In the latter context we note that in a
fermionic quantum fluid there potentially (depending on the values of the Landau Fermi-liquid parameters) are
many other zero modes that also display $\omega\sim k$ scaling, in addition to the shear velocity. 
These can change the prefactor of the singularity, but not the scaling behavior. Also, in a charged Fermi liquid
(e.g., conduction electrons in metals) the first-sound mode turns into the massive plasmon, so our 
approximation of neglecting pressure fluctuations is valid {\em a fortiori}.

%

\vskip 13cm

\onecolumngrid
\begin{center}
\newpage
{\bf SUPPLEMENTAL MATERIAL}
\end{center}

For the purpose of this Supplemental Material section we use units such that $\hbar = \kB = 1$.

\bigskip
\begin{center}
{\bf Fluctuating-Force Correlations}
\end{center}

For the explicit determination of correlation functions one needs the correlations of the fluctuating forces
${\hat P}_{\perp}$ and ${\hat Q}$ in Eqs.~(8). They follow from the correlations of a more general
Langevin operator that were determined in Ref.~\onlinecite{Kirkpatrick_Belitz_2022S}. ${\hat Q}$ is
related to the fluctuating heat current denoted by $\hat{\bm q}_{\text L}$ in that reference by
\bSe
c_p\,{\hat Q}({\bm k},\omega) =  - i{\bm k}\cdot{\hat{\bm q}}_{\text{L}}({\bm k},\omega)\ ,
\label{eq:S1}
\eSe
where $c_p$ is the specific heat at constant pressure, and ${\hat P}_{\perp}$ is related to the fluctuating stress
tensor ${\hat\tau}_{\text{L}}$  in the same reference by
\bSe
{\hat P}_{\perp}({\bm k},\omega) = \frac{-i}{\rho}\,{\hat k}_{\perp}^i k^j ({\hat\tau}_{\text{L}})_{ij}({\bm k},\omega) \ .
\label{eq:S2}
\eSe
The force correlations are obtained from Eqs.~(3.24) in Ref.~\onlinecite{Kirkpatrick_Belitz_2022S}. One obtains
\bSe
\frac{1}{2}\left\langle\left[{\hat Q}({\bm k}_1,\omega_1),{\hat Q}({\bm k}_2,\omega_2)\right]_{\pm}\right\rangle = 
     2\pi \delta(\omega_1 + \omega_2)\,V\delta_{{\bm k}_1,-{\bm k}_2}\,\frac{D_T}{c_p}\,{\bm k}_1^2 \, \omega_1 T 
     c_{\pm}(\omega_1/2T)\ ,
\label{eq:S3}
\eSe
\bSe
\frac{1}{2}\left\langle\left[{\hat P}_{\perp}({\bm k}_1,\omega_1),{\hat P}_{\perp}({\bm k}_2,\omega_2)\right]_{\pm}\right\rangle = 
     -2\pi \delta(\omega_1 + \omega_2)\,V\delta_{{\bm k}_1,-{\bm k}_2}\,\frac{\nu}{\rho}\,k_1^2 \, \omega_1  
   c_{\pm}(\omega_1/2T)   \ ,
\label{eq:S4}
\eSe
where
\bSe
c_{\pm}(\Omega) =  \begin{cases}       1                             & \text{for} \quad +  \\
                                \coth\Omega & \text{for} \quad -                                   
     \end{cases} 
     \label{eq:S5}
\eSe
The cross correlations vanish,
\bSe
\left\langle\left[{\hat Q}({\bm k}_1,\omega_1), {\hat P}_{\perp}({\bm k}_2,\omega_2)\right]_{\pm}\right\rangle = 0\ .
\label{eq:S6}
\eSe
With Eqs.~(S3) - (S6) the correlation functions in the hydrodynamic regime, Eqs.~(5), (6) in the main text, are
easily determined.

\begin{center}
{\bf Correlation Functions in the Collisionless Regime}
\end{center}

\begin{center}
{\it Kinetic Theory}
\end{center}

For the correlation functions in the collisionless regime one needs to utlilize kinetic theory. Let 
\bSe
f_{\bm p}^{\text{eq}} = \frac{1}{e^{(\epsilon_p - \mu)/T} + 1}
\label{eq:S7}
\eSe
be the equilibrium Fermi-Dirac distribution, with $\epsilon_p$ the single-particle energy and $\mu$ the chemical potential,
and let ${\hat f}({\bm p},{\bm x},t)$ be the operator-valued single-particle 
distribution function which, in the collisionless regime, is governed by
a fluctuating Boltzmann-Landau kinetic equation \cite{Landau_Lifshitz_IX_1991, Kirkpatrick_Belitz_2022S}
\bSe
\partial_t {\hat f}({\bm p},{\bm x},t) + {\bm v}_{\bm p}\cdot{\bm\nabla}_{\bm x}{\hat f}({\bm p},{\bm x},t) = w({\bm p})\,{\hat F}_{\text{L}}({\bm p},{\bm x},t)\ .
\label{eq:S8}
\eSe
Here
\bSe
w({\bm p}) = -\partial f_{\bm p}^{\text{eq}}/\partial\epsilon_p = \frac{1}{4T\cosh^2((\epsilon_p - \mu)/2T)}\ ,
\label{eq:S9}
\eSe
and ${\hat F}_{\text{L}}$ is a fluctuating-force operator whose correlations were given in Sec.~II.C of Ref.~\onlinecite{Kirkpatrick_Belitz_2022S}.
Now consider the local equilibrium distribution
\bSe
f_{\bm p}^{(0)}({\bm x},t) = \frac{1}{\exp\left[\left(\frac{({\bm p} - m{\bm u}({\bm x},t))^2}{2m} - \mu({\bm x},t)\right)/T({\bm x},t)\right] + 1}\ ,
\label{eq:S.9.1}
\eSe
with ${\bm u}$, $\mu$, and $T$ the local fluid velocity, chemical potential, and temperature, respectively \cite{Chapman_Cowling_1970},
and write
\bSe
{\hat f}({\bm p},{\bm x},t) = f_{\bm p}^{(0)}({\bm x},t) + w({\bm p})\,{\hat\phi}({\bm p},{\bm x},t)\ .
\label{eq:S10}
\eSe
In the presence of an externally imposed temperature gradient ${\bm\nabla} T$ the streaming term in Eq.~(S8) acting on the local
equilibrium distribution produces a term proportional to ${\bm\nabla} T$. We
neglect pressure fluctuations as we did in the hydrodynamic regime, and keep only the crucial coupling of ${\bm\nabla}T$ to the
shear velocity, ignoring all other effects of the temperature gradient. It is convenient to define functions
\bSe
a_5({\bm p}) = \epsilon_p - \mu + T (\partial\mu/\partial T)_{N,V}
\label{eq:S11}
\eSe
and
\bSe
a_s({\bm p}) = a_5({\bm p}) - (T/n)(\partial p/\partial T)_{N,V} = \epsilon_p - \mu - sT/n
\label{eq:S12}
\eSe
with $s/n$ the averaged entropy per particle. Then we obtain for the linearized version of Eq.~(S8)
\bSe
\left(\partial_t + {\bm v}_{\bm p}\cdot{\bm\nabla}_{\bm x}\right) {\hat\phi}({\bm p},{\bm x},t) = {\hat F}_{\text{L}}({\bm p},{\bm x},t)
 - {\hat u}_{\perp}({\bm x},t)\,\frac{\perpgradT}{T}\,a_s({\bm p})\ .
\label{eq:S13}
\eSe
This generalizes Eq.~(2.6b) in Ref.~\onlinecite{Kirkpatrick_Belitz_2022S} to the
situation of a constant temperature gradient while neglecting the Fermi-liquid interactions. Performing Fourier transforms in
space and time, we now have
\bSe
{\hat\phi}({\bm p},{\bm k},\omega) = G_0({\bm p},{\bm k},\omega) \biggl[{\hat F}_{\text{L}}({\bm p},{\bm k},\omega) 
 - \hat{u}_{\perp}({\bm k},\omega)\, \frac{\perpgradT}{T}\,a_s({\bm p})\biggr] \ , 
\label{eq:S14}
\eSe
with
\bSe
G_0({\bm p},{\bm k},\omega) = \frac{i}{\omega - {\bm k}\cdot{\bm p}/m + i0}
\label{eq:S15}
\eSe
a Green function.

\begin{center}
{\it Correlation Functions}
\end{center}

Temperature fluctuations are defined as linear combinations of fluctuations of the internal energy density $e$ and
the number density $n$,
\bSe
\delta T = \left(\frac{\partial T}{\partial e}\right)_{n,V} \delta e + \left(\frac{\partial T}{\partial n}\right)_{e,V} \delta n
\label{eq:S16}
\eSe
This is true classically, and it remains true if the classical fluctuations are replaced by their operator-value quantum mechanical counterparts.
In terms of the function $\hat\phi$ this implies
\bSe
\delta {\hat T}({\bm k},\omega) = \frac{1}{c_V}\,\frac{1}{V}\sum_{\bm p} a_5({\bm p})\,w({\bm p})\,{\hat\phi}({\bm p},{\bm k},\omega)\ .
\label{eq:S17}
\eSe
Substituting (\ref{eq:S14}) this becomes
\bSe
\delta{\hat T}({\bm k},\omega) = \frac{1}{c_V}\,\frac{1}{V}\sum_{\bm p} w({\bm p})\,a_5({\bm p})\,G_0({\bm p},{\bm k},\omega)\,{\hat F}_{\text{L}}({\bm p},{\bm k},\omega)
  - \perpgradT\,{\hat u}_{\perp}({\bm k},\omega)\,\tau({\bm k},\omega) \ ,
\label{eq:S18}
\eSe
where
\bSe
\tau({\bm k},\omega) = \frac{1}{c_V T}\,\frac{1}{V}\sum_{\bm p} w({\bm p})\,a_5({\bm p})\,a_s({\bm p})\,G_0({\bm p},{\bm k},\omega)\ .
\label{eq:S19}
\eSe
In the low-temperature limit this becomes
\bSe
\tau({\bm k},\omega) = \frac{\NF T}{c_V}\,\frac{\pi^2}{6}\,\frac{-i}{\vF k}\,\log\left(\frac{1 - \omega/\vF k - i0}{-1 - \omega/\vF k - i0}\right) 
+ O(T^3)\ .
\label{eq:S20}
\eSe
We see that the equilibrium part of the temperature fluctuations is given by the fluctuating force, whereas the non-equilibrium part is
given by the fluctuations of the shear velocity. The latter are given by
\bSe
{\hat u}_{\perp}({\bm k},\omega) = \frac{1}{\rho}\,\frac{1}{V} \sum_{\bm p} ({\hat{\bm k}}_{\perp}\cdot{\bm p})\,w({\bm p})\,{\hat\phi}({\bm p},{\bm k},\omega)\ .
\label{eq:S21}
\eSe
Using (\ref{eq:S14}) again, and the fluctuating force correlations from Ref.~\onlinecite{Kirkpatrick_Belitz_2022S}, we find, to leading order as $T\to 0$,
\bSe
\frac{1}{2} \Big\langle \left[{\hat u}_{\perp}({\bm k}_1,\omega_1), {\hat u}_{\perp}({\bm k}_2,\omega_2)\right]_{\pm}\Big\rangle  
               = 2\pi \delta(\omega_1+\omega_2)\,
V\,\delta_{{\bm k}_1,-{\bm k}_2}\,\frac{\pi}{\rho^2}\,\omega_1\,\frac{1}{V}\sum_{\bm p} w({\bm p})\,
     \left(\hat{\bm k}_{1\perp}\cdot{\bm p}\right)^2 \delta(\omega_1 - {\bm k}_1\cdot{\bm p}/m)\,c_{\pm}(\omega_1/2T)
\label{eq:S22}
\eSe
with $c_{\pm}$ from (S5). Here we have used the expressions for the fluctuating-force correlations given in Ref.~\onlinecite{Kirkpatrick_Belitz_2022S},
in particular (2.11), (2.18), (2.15c), and (3.7) in that paper. 

We can now assemble the temperature-temperature correlation functions. We find
\begin{equation*}
\chi''_{TT}({\bm k},\omega) = \frac{\pi}{c_V^2}\,\frac{1}{V} \sum_{\bm p} w({\bm p})\,\left(a_5({\bm p})\right)^2 \omega\,\delta(\omega - {\bm k}\cdot{\bm p}/m) \hskip 220pt \mbox{}
\end{equation*}
\vskip -15pt
\bSe
    \qquad +\, \omega\,\frac{\pi}{4}\,\frac{(\gradperpT)^2}{\vF^2 k^2}\,\frac{\kF^2}{\rho^2}\,\frac{1}{V}\sum_{\bm p} w({\bm p}) \left(1 - (\hat{\bm k}\cdot\hat{\bm p})^2\right) 
     \left[\log^2 \left\vert\frac{1 - \hat{\bm k}\cdot\hat{\bm p}}{1 + \hat{\bm k}\cdot\hat{\bm p}}\right\vert + \pi^2\right]\,\delta(\omega - {\bm k}\cdot{\bm p}/m) 
\label{eq:S23}
\eSe
and
\bSe
S^{\text{sym}}_{TT}({\bm k},\omega) = \chi''_{TT}({\bm k},\omega)\, \coth(\omega/2T)
\label{eq:S24}
\eSe
From these expressions one can determine the static correlation functions. At $T=0$, $w({\bm p})$ turns into a delta function and for $\chi_{TT}({\bm k})$ one obtains
obtains Eq. (7) in the main text. 


\begin{thebibliography}{32}
\bibitem[cla()]{classical_limit_footnote}
Note that the commutator correlations still exist in the classical limit, due
  to the factor of $\hbar$ in the definition. Alternatively, they can be
  defined in terms of Poisson brackets.
\end{thebibliography}

\end{document}